\documentclass[12pt]{article}
\usepackage{graphicx}
\usepackage{amsmath}
\usepackage{lipsum}
\usepackage{hyperref}
\usepackage{enumitem}
\usepackage{float}
\usepackage{enumitem}

\title{\textbf{Single Block On}}
\author{
 Paritosh Ranjan \\
  IBM  \\
  \texttt{paranjan@in.ibm.com} \\
  \and
 Surajit Majumder \\
  IBM  \\
  \texttt{surajit.majumder@ibm.com} \\
  \and
 Prodip Roy \\
  IBM  \\
  \texttt{prodipro@in.ibm.com} \\
}
\date{\today}

\begin{document}
\maketitle

\begin{abstract}
In the digital age, individuals increasingly maintain active presences across multiple platforms—ranging from social media and messaging applications to professional and communication tools. However, the current model for managing user-level privacy and abuse is siloed, requiring users to block undesirable contacts independently on each platform. This paper introduces Single Block On (SBO)—a unified and interoperable system enabling users to block an individual once and have that block propagated across all integrated applications. SBO operates via identity-based matching rules, utilizing configurable levels of identifier similarity, and interfaces with systems through standardized protocols such as SSO, LDAP, or direct REST integration. A novel Contact Rule Markup Language (CRML) facilitates consistent policy sharing across systems. The proposed solution increases user safety, enhances digital well-being, and sets a precedent for interoperable privacy enforcement.

\end{abstract}

\section{Introduction}

The increasing fragmentation of digital ecosystems has given rise to a critical gap in user control and safety management—specifically, in the ability to block other users across platforms. Today, if a user wishes to block a harassing or unwanted individual, they must independently initiate blocks across every application they use: from social media (e.g., Facebook, Instagram) and messaging apps (e.g., WhatsApp) to email providers and collaboration tools. This manual process is inefficient, error-prone, and fails to provide users with holistic protection.

This paper proposes a solution that mirrors the philosophy of Single Sign-On (SSO), where a user logs in once to gain access across multiple systems. We call this Single Block On (SBO)—a mechanism by which blocking a contact in one centralized system automatically extends that block across all user-authorized applications.

To the best of our knowledge, no existing invention or standard provides a user-centric, identity-based, and interoperable block list mechanism. We present SBO as a novel system that fills this technological and social gap.
 
\section{Brief Description of the Invention}

This invention proposes that if anyone blocks one or more user’s email/user id/username/phone number/profile image or any other identifier in their “Single Block On” facility, then those users having those blocked user ids or other identifiers would be blocked on all digital platforms in which the blocking user has integrated their “Single Block On” facility.
This invention will make it easier to block users across applications once identified in any application. It will not only save time, but also ensure that any unwanted user cannot continue to appear unblocked on any application using any of the similar identifiers.

\section{Reduction to Practice}

The proposed Single Block On system has been conceptually designed and is capable of practical realization via current web and authentication technologies. The system architecture includes the following components:

\subsection{SBO Providers}
Entities that host and manage block lists. Examples include sbo.aws.com, sbo.ibm.com, or sbo.free.com. Users create SBO accounts with these providers and define “Block Lists” containing unwanted contacts.

\subsection{User-Created Block Lists}
Users can define lists based on identifiers such as:
\begin{enumerate}
    \item Email addresses
    \item Phone numbers
    \item Usernames
    \item Profile pictures
    \item Biographical information
    \item Other contact metadata
\end{enumerate}

Users configure matching rules with specified strictness levels: Strict, Medium, or Lenient.

\subsection{Contact Rule Markup Language (CRML)}
A structured data format (XML/JSON) to describe block lists and associated matching rules. CRML enables communication between SBO providers and client applications.

\subsection {Integration Methods}
Applications can consume SBO data through various methods:

\begin{enumerate}
    \item Via SSO providers (e.g., OAuth)
    \item Via LDAP servers
    \item Direct integration through REST APIs
    \item Manual provision of SBO credentials during app login
    \item Priority rules allow for multiple SBO integration strategies.
\end{enumerate}

\subsection {Rule-Based Matching Engine}
Applications use the identifiers and rules specified in CRML to match and block corresponding user profiles. Matching can include fuzzy matching and logical operations (AND, OR, etc.).

\subsection {Real-Time Enforcement}

Block lists can be refreshed:

\begin{enumerate}
    \item Periodically (auto-refresh)
    \item On login
    \item On each request
    \item Manually via user or system triggers
\end{enumerate}
\subsection {Multi-Provider Support}
Users can register with multiple SBO services, with all block lists considered during application enforcement.

This complete pipeline—from identifier registration to real-time enforcement—demonstrates that the proposed invention is not only theoretically sound but also readily implementable with existing technologies.

\subsection{\textbf{Following are the steps of Implementation}}

\begin{itemize}
    \item Free or paid “Single Block On/SBO” providers will be available on the internet. For example sbo.google.com, sbo.ibm.com,  sbo.aws.com, sbo.free.com, sbo.paid.com etc.
    \item A user will create an account with a “Single Block On” provider on web. For example, the created account named “alexandergrahambell” with SBO provider sbo.aws.com
    \item The user will create a block list named “Block List 1” on sbo.aws.com in that account.
    \item In the block list, the user will add the contacts which should be blocked. 
    \item The user will provider identifiers whose similarity should be considered for each added contact. 

    \item The identifiers can be textual, or image based.
    \item Examples of identifiers are:
            \begin{enumerate}
            \item Full Name
            \item Email Id
            \item  Phone Number
            \item  Profile Image
            \item  Username i.e., the username name used by the Contact in            applications
            \item  Gender
            \item  Age
            \item  Location
            \item  Biodata
            \item  A mix of these identifiers
            \end{enumerate}
    \item The system will come configured with the default list of selected contact identifiers. The user can override that list and provide its own list of identifiers.
    \item The user will also provide the level of strictness of similarity to enforce while matching Contacts i.e., Strict, Medium, or Lenient.
            \begin{enumerate}
            \item For Strict matching, the values of the identifiers should be equal or near equal textually.
            \item For Medium matching, the values of the identifiers can differ by few characters.
            \item  For Lenient matching, the values of the identifiers can differ by more characters than in medium matching.
            \end{enumerate}

    \item The user can configure the SBO account and the BLOCK LIST at following levels:
            \begin{enumerate}
            \item Configure/Integrate the SBO with the SSO provider: 
            When the application will authenticate via SSO, then the application will fetch the SBO and block list details from the SSO provider for the authenticated account. The user will authorize the SSO provider to authenticate to the configured SBO and extract the details of the configured Block List.
            
            \item Configure/Integrate the SBO with the LDAP provider:
            When the application will authenticate via LDAP, then the application will fetch the SBO and block list details from the LDAP server for the authenticated account. The user will authorize the LDAP server to authenticate to the configured SBO and extract the details of the configured Block List.
            
            \item Configure/Integrate the SBO directly in the application:
            The user can configure/integrate the SBO details directly in the application. The user can authorize the application to authenticate to the SBO provider and fetch the details of the block list.
            \item Provide the SBO details to the app at the time of logging in:
            The user can provide the SBO provider details and the Block List name at the time of logging into the application.
            \item The user can configure the priority of each type of SBO integration. 
            If the user configures the SBO using multiple methods, then the available/configured SBO integration method having highest priority will override all other SBO integrations provided by the user.
            \end{enumerate}    

    \item The system will provide multiple methods to match Contacts. The application owners can also provider their own custom methods for Contact Matching.
    \textbf{Rule Based Matching}
            \begin{enumerate}
            \item The user will select the profile that the user wants to block.
            \item The system will identify all contact identifier entities on the profile e.g., Name, Email, Data of Birth, Gender, Address, Education etc.
            \item This automatic detection of identifiers of a Contact for the creation of matching rules is visualized in the image below.
            \begin{figure}
                \centering
                \includegraphics[width=0.5\linewidth]{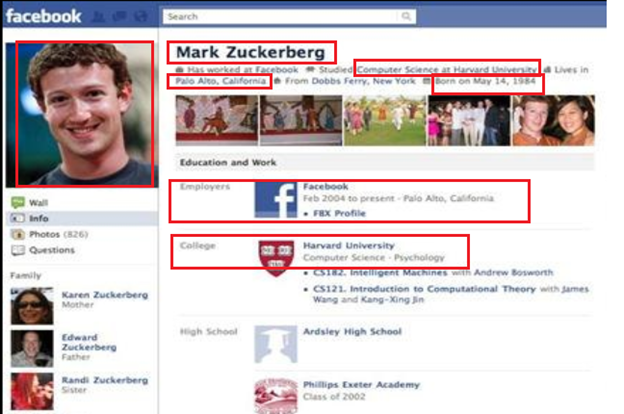}
                \caption{Contact-Identifiers-detected-by-the-LLM}
                \label{fig:Contact-Identifiers-detected-by-the-LLM}
            \end{figure}
            \newpage
            \item The user will be able to select these identifiers to create MATCHING RULE at the block list level for the contact to match.
            \item Once the user selects any matching identifier the user will get options like MATCHES, EQUALS, GREATER THAN, FUZZYMATCHES, AND, OR etc.
            \item Using these options, the user will be able to use the UI to create matching rules which will internally look like following.
            E.g., 
                (Full Name MATCHES AND Phone Number MATCHES)
                OR
                (Photograph MATCHES AND Gender MATCHES AND Location MATCHES)
                OR
                (Bio MATCHES AND Email Id Matches)
                OR 
                (Username MATCHES AND Bio FUZZYMATCHES)

            \item The system will also come with default matching rules.
            \item The default MATCHING RULE will be provided by the SBO provider. The user can override that rule.
            \end{enumerate}    
    \item The application will fetch the Contact Details and Matching Rule’s details from the SBO provider for the Block List which was found to have the highest priority.
    \item The Contact Details and the Matching method’s details will be shared by the SBO provider to the application via REST in a structured format.
    \item This structured format can be called CRML i.e.,” Contact Rule Markup Language”(like SAML). It can be a structured XML or JSON.
    \item CRML will contain information about the contacts and the contact matching rule’s details.
    \item When any user will log in, then the application will iterate over the Contact List received from the SBO provider, and match the contacts in the list with all the Profiles interacting with the user using the MATCHING method’s details received from the SBO provider via CRML.
    \item The application will block access and visibility to all the users whose profiles will match the retrieved list.
    \item When any blocked user will login, then the application will fetch the users from all integrated SBO providers who have blocked this user in one or more SBO providers.
    \item The application will disable/block access of the blocked user to the users/user account who have blocked this user.
    \item If the user updates the Block List in the SBO, then the new block list can be refreshed in the application by:
            \begin{enumerate}
            \item Refreshing Block List automatically after every T seconds/minutes etc.
            
            \item Triggering refresh of cached block list manually in the application.
            
            \item Logging out and logging in again as every login can be configured to extract the fresh block list.
            \item Every new request to the application checks for update in the block list.
            \end{enumerate}    
    \item The user can also configure multiple SBO providers and multiple Block Lists in each SSO account. The application will apply to all the block lists received from all the configured SBO accounts.

\end{itemize}

\begin{figure}
    \centering
    \includegraphics[width=0.9\linewidth]{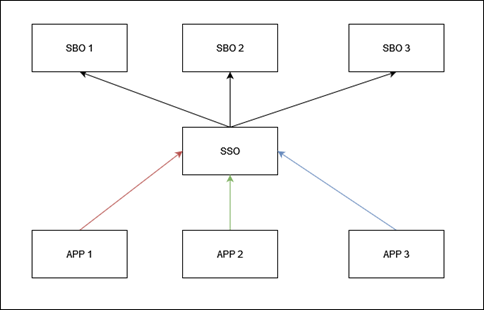}
    \caption{Application-Integrating-with-SBO-via-SSO}
    \label{fig:Application-Integrating-with-SBO-via-SSO}
\end{figure}

\begin{figure}
    \centering
    \includegraphics[width=0.9\linewidth]{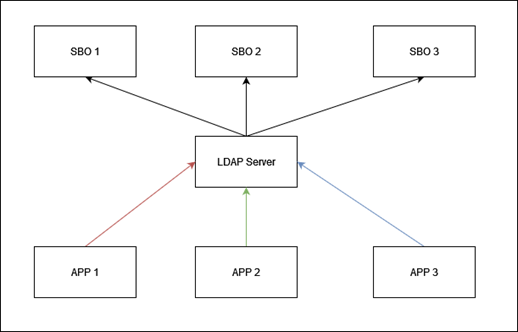}
    \caption{Application-Integrating-with-SBO-via-LDAP}
    \label{fig:Application-Integrating-with-SBO-via-LDAP}
\end{figure}

\begin{figure}
    \centering
    \includegraphics[width=0.9\linewidth]{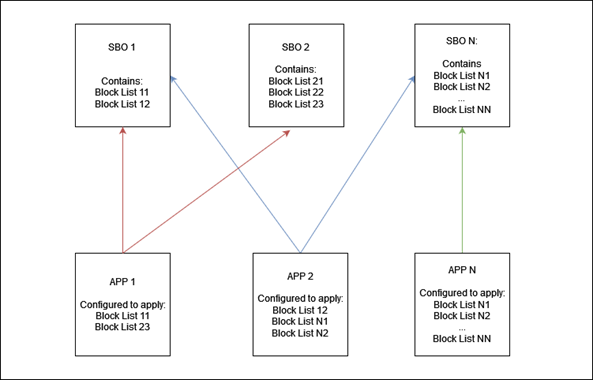}
    \caption{Direct-Integration-of-SBO-with-Application}
    \label{fig:Direct-Integration-of-SBO-with-Application}
\end{figure}

\newpage
\section{Advantages of the Invention}
The Single Block On invention provides several advantages:

\begin{enumerate}
    \item Unified Privacy Management : 
Users can centrally manage all block actions from one interface rather than repeating the same action across multiple platforms.
    \item Cross-Platform Consistency : 
Once blocked in SBO, a contact is effectively blocked everywhere—on email, messaging, social, professional, and collaborative platforms that the user connects.
    \item Time and Cognitive Load Reduction :
Instead of remembering to block an individual across 10 different systems, the user performs the action once.
    \item Interoperable and Vendor-Agnostic :
SBO is designed to integrate with multiple identity systems and is platform-agnostic, allowing independent developers and organizations to adopt it easily.
    \item Standardization Through CRML :
CRML serves as a de facto schema standard to facilitate interoperability between SBO services and applications.
    \item Configurable Identity Matching :
Supports multiple identity types (email, phone, profile picture, etc.) and allows for fuzzy matching and customizable rule logic.
    \item Improved User Safety :
Harassment, spam, or abuse mitigation becomes more robust, especially for vulnerable populations or high-profile individuals.
\end{enumerate}

\section{Conclusion}

This paper presents Single Block On (SBO), a new paradigm in digital user safety and privacy management. By extending the metaphor of Single Sign-On to blocking functionality, SBO allows users to block malicious or unwanted contacts across multiple platforms via a single action. Through the use of rule-based matching, multi-identifier configuration, real-time enforcement mechanisms, and the introduction of the Contact Rule Markup Language (CRML), SBO is both a practical and innovative system capable of revolutionizing digital privacy practices.

The proposed system is ripe for implementation using existing identity and application infrastructure. Future work includes defining the CRML standard through open consortia, developing open-source reference implementations, and promoting industry-wide adoption. SBO represents a meaningful advancement in how users manage their online presence and interactions in a decentralized digital world.

\section{Acknowledgment}

We would like to express our sincere gratitude to all individuals and organizations who have contributed to the success of this research. We acknowledge the invaluable support from the IBM team, whose resources and expertise have greatly enhanced this project.
Special thanks to Prodip Roy (Program Manager IBM) for their insightful feedback, guidance, and encouragement throughout the development of this work.

\section{References}
\renewcommand\refname{}

\end{document}